# Une brève introduction à la matière molle

Fabrice Cousin[1], Isabelle Grillo[2], Jacques Jestin[1] et Julian Oberdisse[3]

[1] Laboratoire Léon Brillouin, CEA-CNRS UMR 12, CEA Saclay, 91191 Gif sur Yvette
[2] Institut Laue Langevin, DS-LSS, 6 rue Jules Horowitz, B.P. 156, 38042 Grenoble Cedex 9
[3] Université Montpellier II, Laboratoire des Colloïdes, Verres, et Nanomatériaux, UMR CNRS 5587, Place E. Bataillon, F-34095 Montpellier Cedex

Le propos de ce cours n'est pas de faire un descriptif exhaustif de la matière molle, il existe un certain nombre d'ouvrages que le lecteur désireux d'en savoir plus pourra consulter [1-5]. L'objectif est de faciliter la lecture du présent ouvrage en présentant les systèmes, et les questions physiques sous-jacentes, afin de montrer pourquoi la diffusion de neutrons est importante dans le domaine. Le lecteur pourra ainsi faire le lien entre les cours théoriques sur les techniques neutroniques et les cours d'illustration d'exemples.

Définir ce qu'est la *matière molle* n'est pas aisé car on désigne à travers ce terme des systèmes physico-chimiques très variés, qu'on appelle également souvent *fluides complexes*, qui forment un domaine dont les limites sont floues[*]. L'expression est semble-t-il d'origine française, puisqu'elle aurait été inventé par la physicienne Madeleine Veyssié dans les années 70 au Laboratoire de Physique des Solides d'Orsay, *'…pour désigner tout ce qui va des matières plastiques aux bulles de savon, en passant par les gels, les élastomères, les cristaux liquides, les crèmes cosmétiques, les boues, les pâtes céramiques, etc..'* [6]. Si cette expression n'était peut-être qu'une simple boutade à l'origine, elle a fait florès au point qu'un grand journal de la Royal Society of Chemistry traitant du domaine, né en 2005, *Soft Matter* [7], a pris pour nom sa traduction littérale en langue anglaise (les anglo-saxons utilisent également le terme 'Soft Condensed Matter'). Le point commun à tous les systèmes de la matière molle, qui peut servir de définition, est que les énergies d'interaction mises en jeu entre objets (liaisons H, interactions de Van der Waals, etc…) sont comparables à l'énergie thermique kT à température ambiante. Les effets enthalpiques étant du même ordre de grandeur que les effets entropiques, les systèmes sont susceptibles de se réorganiser fortement sous l'effet de variations faibles de l'environnement (température, pression, concentration) ou de faibles sollicitations extérieures (contrainte mécanique, champ électrique, champ magnétique, etc..). Du fait du grand nombre d'échelles mises en jeu (énergétiques, spatiales, temporelles), la *physique de la matière molle* est donc intermédiaire entre la physique des liquides et la physique des solides

Les objets caractéristiques constitutifs des systèmes de la matière molle sont les polymères, les colloïdes, les tensioactifs, les cristaux liquides, les protéines ou plus généralement les biopolymères Nous résumons dans les différents encadrés paragraphes qui suivent les caractéristiques importantes des polymères, des colloïdes et des tensioactifs. Ils forment les 'briques élémentaires' d'assemblages ou de matériaux ayant un vaste champ d'applications industrielles dans des domaines tels que l'industrie des plastiques, des cosmétiques, de l'agroalimentaire, la pharmacologie… Le domaine est en plein essor, en particulier parce que les progrès réalisés par les chimistes lors de ces dernières années sur la synthèse des objets (contrôle de la taille et de la morphologie de nanoparticules, click-chemistry, etc..) permettent de réaliser en pratique une infinité d'architectures associant deux (ou

---

[*] Il n'existe d'ailleurs pas de définition exacte du terme dans la nomenclature IUPAC de *l'Union internationale de chimie pure et appliquée.*

plus..) types d'objets afin de combiner les propriétés remarquables des objets individuels (mécaniques, optiques, magnétiques, etc…). Il y a par exemple à l'heure actuelle des recherches très actives sur le développement de matériaux appelés *intelligents*, capables de changer spontanément de forme ou de couleur sous l'effet d'un stimulus externe tel que le pH ou la température... Les systèmes formés d'objets organisés en 2D ont également un rôle technologique très important puisqu'ils permettent de moduler des propriétés telles que la tension de surface, le mouillage, l'adhésion…

Les systèmes de la matière molle présentent les points communs suivants:
- Les échelles de taille caractéristiques des objets qui les constituent varient du nanomètre à la centaine de nanomètres.
- Dans un système donné, il est extrêmement rare, à l'exception des protéines ou de certains biopolymères comme l'ADN, que tous les objets aient exactement la même forme et/ou la même taille. Cette *polydispersité* peut modifier fortement les propriétés physiques des systèmes. A l'échelle des assemblages, puisqu'ils sont non covalents, les fluctuations thermiques peuvent induire des fluctuations de taille importante, aussi bien temporelles que d'ensemble.
- Les objets ayant une petite taille, ils ont de grandes surfaces spécifiques. Les effets surfaciques et interfaciaux sont donc souvent très importants et deviennent parfois prépondérants par rapports aux effets de volume.
- L'importance des effets entropiques sur la physique des systèmes peut modifier fortement leurs propriétés lorsque l'on change leur dimensionnalité (couches minces 2D) ou que l'on confine les objets.
- Du fait de la complexité des systèmes, l'exploration de l'espace des phases se fait souvent en plusieurs étapes distinctes et peut varier très fortement entre deux systèmes très proches du point de vue de la composition. La mise à l'équilibre des systèmes présente alors plusieurs temps caractéristiques et peut varier sur plusieurs décades temporelles pour atteindre des temps extrêmement lents, supérieurs à la seconde, dans certains cas. Puisque les effets enthalpiques sont du même ordre que les effets entropiques, il arrive par ailleurs que les systèmes restent cinétiquement piégés dans des minima lors de l'exploration de l'espace des phases et restent hors d'équilibre. C'est le cas suspensions colloïdales, qui vitrifient à haute concentration et servent de systèmes modèles pour la compréhension des mécanismes de la transition vitreuse.

## 1. Les polymères

Les polymères ou macromolécules sont des molécules faites d'un grand nombre d'unités de répétition, ou monomères. Leur diversité chimique – homopolymères faits d'un seul monomère, des copolymères statistiques ou à bloc, etc…en partant d'une grande variété de monomères existante – et topologique – des molécules linéaires, branchés, en étoile, en anneau, … – semble plus ou moins illimitée. Ajoutons des fonctions ou des propriétés particulières à cette diversité (caractère hydrophobe/hydrophile, agrégation ou association, polyélectrolyte, possibilités de complexation avec des ions ou des particules, …), sans parler de leurs applications (de la solution polymère au matériau en volume) et il apparait clairement que le grand nombre de *conformations* possibles des macromolécules fournira des sujets de thèse encore pendant longtemps.

Une macromolécule possédant par définition un grand nombre de sous-unités, il s'agit d'objets nécessitant intrinsèquement une approche statistique pour décrire leurs conformations. Le cas le plus simple est celui d'une chaîne linéaire idéale, comme on peut l'observer en solution dite 'Θ', ou en fondu. La conformation s'apparente à une pelote statistique, dont la taille typique est donnée par le rayon de giration :

$$R_g = a/\sqrt{6} \ N^{1/2}$$

où a est la longueur de l'unité de répétition, et N le nombre de motifs. Il est possible de mesurer $R_g$ dans une expérience de diffusion de neutrons aux petits angles, et c'est précisément ce type d'application où les neutrons sont performants, notamment grâce à la substitution isotopique permettant de rendre visible certaines molécules uniquement. Une conséquence directe de la dépendance du $R_g$ en $N^{1/2}$ est une dépendance angulaire de l'intensité diffusée suivant une loi d'échelle caractéristique, des déviations de l'idéalité se manifestant par des lois d'échelle différentes. Par ce biais, la diffusion de neutrons a fortement interagi avec les 'Scaling concepts in polymer physics' popularisés par P.G. de Gennes [5]. Bien entendu, la mesure des conformations va aujourd'hui bien au-delà de ce cas très simple, et différents exemples ont été traités dans le cadre des cours donnés aux JDN17, reproduits dans ce volume.

## 2. Les dispersions colloïdales

Un système colloïdal est une dispersion dans un milieu porteur d'objets soumis à une agitation Brownienne suffisamment importante pour ne pas sédimenter. Il existe des dispersions de types variés telles que les suspensions colloïdales (dispersions de particules solides dans un liquide), les fumées (particules solides dans un gaz), les mousses (objets gazeux dans un liquide), ou les émulsions (objets liquides dans un liquide) [1]. Les particules, appelées colloïdes, ont une taille comprise entre 1 nm et 500 nm. La limite inférieure est imposée par l'obligation de pouvoir considérer la suspension comme un système pour lequel les échelles de taille entre les particules et les molécules du milieu porteur sont très différentes. La limite supérieure est imposée par l'obligation de pouvoir considérer le mouvement des particules comme complètement Brownien et non perturbé par des effets de gravitation ou d'interactions hydrodynamiques.

Les particules, d'origine naturelle ou synthétique, peuvent avoir des tailles et des morphologies variées (billes, sphères, disques, bâtons, cubes, haltères, etc…) ainsi que des propriétés physiques intéressantes (optiques, catalytiques, magnétiques…) provenant soit des propriétés intrinsèques des matériaux, soit d'effets quantiques liées à leur taille réduite (résonance plasmon, quantum dots…).

Dans toute suspension, la présence des forces attractives de Van der Waals tend à agréger les particules. Il est donc nécessaire de contrebalancer ces interactions attractives, auxquelles peuvent s'ajouter d'autres forces attractives telles que les forces de déplétion ou les forces dipolaires magnétiques, par des interactions répulsives pour maintenir la dispersion des particules. Ces forces répulsives peuvent être de nature entropique dans le cas de l'adsorption ou du greffage de polymères, ou, dans le cas très courant des suspensions de particules chargées dans des solvants polaires, de nature électrostatique. C'est donc le potentiel interparticulaire, qui est en fait une variation d'énergie libre, somme des interactions attractives et répulsives, entropiques et enthalpiques, qui détermine la structure des dispersions à l'échelle locale. Le potentiel le plus connu, qui sert de base à la théorie des colloïdes chargés depuis les années 1940, est le potentiel DLVO (du nom des russes Derjaguin et Landau et des hollandais Vervey et Overbeek) qui décrit les répulsions électrostatiques sous la forme d'un potentiel de Yukawa ayant comme portée caractéristique la longueur de Debye [4].

Il est fondamental de pouvoir mesurer expérimentalement la structure locale des dispersions, car elle détermine *in fine* la plupart des propriétés thermodynamiques macroscopiques des dispersions. C'est pourquoi la diffusion de rayonnement joue un rôle important dans le monde des colloïdes, car elle permet la détermination de cette structure sur des ensembles statistiques représentatifs, ainsi que la forme et la taille des colloïdes.

## 3. Les tensioactifs

Les molécules tensioactives (surfactant en anglais pour **surf**ace **act**ive **ag**ent) se caractérisent par une tête polaire hydrophile et un corps hydrophobe. Ce caractère amphiphile leur confère des propriétés d'auto-association exceptionnelles lorsqu'elles se trouvent en solution. Elles présentent également une affinité particulière pour les interfaces (liquide/liquide, solide/liquide, gaz/liquide) dont elles abaissent l'énergie libre et la tension de surface. Les tensioactifs sont classés en quatre catégories en fonction de la nature de leur tête polaire. Les cationiques ont une tête polaire chargée positivement et un contre-ion négatif ; inversement, les anioniques ont une tête chargée négativement et un contre-ion positif. Les zwitterioniques (ou amphotères) portent à la fois une charge positive et une charge négative, leur charge globale est neutre. Les non-ioniques ne portent aucune charge mais possèdent un groupement très polaire.

Les tensioactifs ont des nombreuses applications industrielles (agro-alimentaire, textile, pétrole et des matières plastiques, engrais, cosmétiques et produits d'entretien, la pharmacie,…). Il est donc capital d'un point de vue fondamental de comprendre les moteurs de l'organisation de ces molécules, les structures et interactions et la stabilité des objets ainsi formés, afin de relier les propriétés physiques macroscopiques à l'organisation nano- et mésoscopique.

Les propriétés d'adsorption aux surfaces et d'auto-assemblage en milieu aqueux sont principalement gouvernées par les forces hydrophobes, c'est-à-dire par l'expulsion des chaines aliphatiques de l'eau. Ainsi, en fonction de la concentration, de la température ou de la force ionique, chaque tensioactif présente une organisation spécifique. Néanmoins, on peut dégager des comportements généraux. L'agrégation apparaît au-delà d'une concentration appelée « concentration critique d'agrégation » (cac) ou « concentration micellaire critique» (cmc) en fonction du type d'agrégats formés. Au-delà de cette concentration, des agrégats d'une centaine de molécules (micelles) ayant une taille caractéristique entre 20 et 60 Å sont en équilibre thermodynamique avec des monomères en solution. En milieu dilué, la DNPA est une méthode puissante pour déterminer la forme et la taille de ces agrégats. Lorsque la concentration en tensioactif augmente, le nombre de micelles augmente également et elles se rapprochent les unes des autres. Typiquement au-delà de 30% en masse, les forces répulsives (hydratation et électrostatiques) forcent le système à se réorganiser en changeant de forme et de taille. Une succession de phases lyotropes (si induite par le solvant) est généralement observée dans le domaine concentré, comme présenté sur la Figure 1. Si ces structures gardent une dimension de l'ordre de quelques dizaines d'Angströms, on observe des structures à grandes échelles, jusqu'au dixième du micromètre, qui présentent souvent des motifs de répétition, d'où le terme de « cristal liquide ». Mesurée sur un large domaine de vecteur d'onde, $(10^{-3} - 0.1 \ \text{Å}^{-1})$, la DNPA met à la fois en évidence la forme et l'organisation de l'échantillon.

Il est possible de comprendre la séquence de morphologies des agrégats formés à partir du paramètre d'empilement $p$, introduit par Tanford [8] puis Israeslachvili *et al* [9], qui relie la géométrie de la molécule à la courbure de l'interface. Il est défini comme $p = \dfrac{v}{a_0 l_c}$ où $v$ est le volume hydrophobe, $l_c$, la longueur de la chaine alkyle et $a_0$ la surface par tête polaire à la surface de l'agrégat. Pour p<1/3, la structure attendue est sphérique ou ellipsoïdale, avec une courbure importante de l'interface; entre 1/3 et 0.5, des agrégats cylindriques se forment ; de 0.5 à 1 des objets plans, comme des membranes et des vésicules sont présents. Pour p>1, on obtient des micelles inverses.

Le volume $v$ est propre à une molécule et est fonction de la longueur, du nombre de chaines, d'éventuelles ramifications. Par contre, la surface $a_0$ dépend de l'hydratation et donc de la température, de la concentration et de la force ionique qui modifient les répulsions électrostatiques entre têtes. $p$ varie donc en fonction des conditions physico-chimiques de la solution, ce qui est à l'origine des différentes phases formées. Pour mieux contrôler la taille et la forme des agrégats, des mélanges de molécules sont utilisés. Ainsi, en variant le rapport molaire entre un tensioactif de paramètre d'empilement faible avec un co-tensioactif dont $p$ est plus élevé, on génère la séquence classique « micelle sphérique, cylindrique, bicouche » [2].

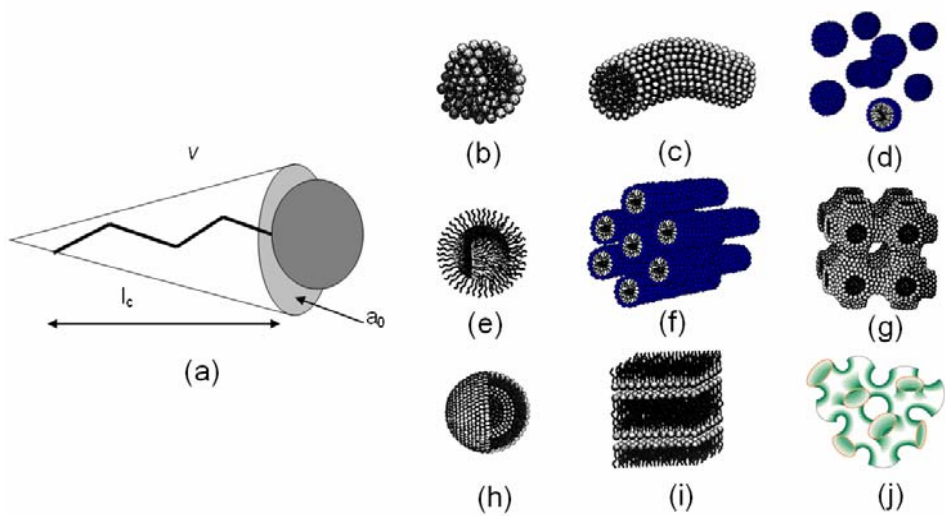

**Figure 1**. Structures des agrégats et phases lyotropes cristallines communément rencontrées dans les systèmes binaires tensioactif/eau. (a) Représentation schématique d'une molécule tensio-active. (b) Micelle sphérique directe (L1) ; (c) Micelle cylindrique (L1) ; (d) Phase cubique (I1); (e) Micelle inverse (L2) ; (f) Phase hexagonale directe(H1); (g) Structure bicontinue cubique (V1); (h) Vésicule ; (i) Phase lamellaire (Lα) ; (j) Structure bicontinue : phase éponge symétrique (L3).

## 4. Pourquoi faire des neutrons ?

Les techniques de diffusion élastiques et inélastiques des neutrons sont bien adaptées à l'étude des systèmes de la matière molle pour plusieurs raisons. Les longueurs d'onde des neutrons incidents, accessibles via un réacteur à fission ou une source à spallation, permettent de sonder des tailles caractéristiques, de 1 à 100 nanomètres, représentant les échelles spatiales pertinentes pour décrire la forme et l'organisation des objets que l'on trouve dans les systèmes classiques de la matière molle. De la même manière, les temps caractéristiques accessibles en diffusion inélastique des neutrons (de la ps à la dizaine de ns) permettent de décrire les processus dynamiques (relaxations, rotations ...) typiques de ces systèmes (voir cours R. Zorn). Les faibles énergies mises en jeux (typiquement le meV) font que le rayonnement neutronique est non destructif pour les échantillons. Par définition neutres, les neutrons n'interagissent pas avec les champs électriques et peuvent donc pénétrer la matière en profondeur (de l'ordre de quelques cm). Ces propriétés en particulier font que dans certains cas, l'utilisation du rayonnement neutronique sera préférable à celui des rayons X. D'une façon générale, la diffusion de rayonnement permet d'obtenir une mesure statistique sur un grand nombre d'atomes (~ $10^{22}$) et donc d'avoir une information moyennée représentative d'un échantillon macroscopique. Ceci constitue un avantage des techniques de diffusion par rapport aux techniques de microscopie, qui certes, donnent une image de l'espace réel mais en un point localisé de l'échantillon et donc pas toujours représentatif de son ensemble (une statistique équivalente obligerait à traiter un nombre considérable d'images). Néanmoins, pour comprendre et décrire les propriétés macroscopiques des systèmes, il est souvent nécessaire de vérifier que les structures définies à l'échelle locale sont conservées ou au contraires s'organisent différemment à plus grande échelle. Pour cela, les techniques de microscopie (qui permettent une résolution jusqu'au µm) mais aussi la diffusion de lumière sont complémentaires des techniques de diffusion des neutrons ou de rayons X. La complémentarité des techniques de mesure est d'une façon générale une notion très importante pour résoudre les problèmes physiques posés, que ce soit en terme d'échelles spatiales (voir cours J.-F. Berret et J. Oberdisse) ou de relations entre structure et dynamique (voir cours S. Lyonnard et E. Dubois). La plupart des constituants en matière molle sont des objets organiques de composition proches (polymères, tensioactifs, protéines...) composés principalement de carbone (C), d'hydrogène (H), d'oxygène (O) et d'azote (N). L'interaction neutron-matière fait que ces éléments sont facilement « visibles » en diffusion de neutrons d'autant plus que les conditions de contraste peuvent être modifiées et améliorées par substitution isotopique. Le cas d'école est bien sur l'hydrogène qui présente un fort contraste neutronique avec le deutérium. Ainsi beaucoup d'expériences de DNPA peuvent être réalisées en remplaçant simplement le solvant hydrogéné (l'eau) par de l'eau lourde pour augmenter le contraste et améliorer significativement les spectres mesurés en volume ou aux interfaces (voir cours G. Fragneto et cours M. Sferrazza). Lorsqu'on s'intéresse à des « systèmes mixtes », constitués de mélanges de plusieurs constituants (deux ou plus...) tels que les mélanges de polymère et de particules inorganiques, les mélanges polyélectrolyte-protéine ou polymères-tensioactifs, on peut utiliser la variation de contraste pour « masquer » sélectivement un des constituants du système et ainsi obtenir un signal de diffusion plus facile à interpréter (voir cours E. Dubois). En terme de contraste, la résolution des structures peut aussi bénéficier de l'utilisation conjointe des techniques de diffusion des neutrons et des rayons X (voir cours J. Combet). Lors d'une approche classique de caractérisation d'un système en DNPA, on cherche dans un premier temps à extraire le facteur de forme des objets constituant ce système. Cela peut être réalisé en travaillant par exemple en solution diluée (ou par écrantage dans les systèmes chargés) de façon à supprimer les interactions entre les objets. Par des analyses simples sur les spectres obtenus, décrites dans le cours de F. Nallet, on pourra assez rapidement obtenir une taille caractéristique (régime de Guinier), une information sur la géométrie (exposants caractéristiques dans le régime intermédiaire) ou un rapport surface/volume (régime de Porod). Dans un second temps, les facteurs de forme peuvent être modélisés sur l'ensemble de la gamme de vecteur d'onde par des formulations analytiques plus élaborées (incluant les effets de polydispersité et de résolution expérimentale, voir le cours de D. Lairez) ou des méthodes numériques (voir le cours de L. Belloni). Une fois les facteurs de formes connus, on s'intéresse aux facteurs de structure et à la nature des interactions dans les systèmes concentrés : on peut obtenir une information directe sur le spectre mesuré, par exemple la distance moyenne entre objets dans un système répulsif à partir de la position du pic ou avoir recours à des méthodes d'analyses plus complexes (voir cours de L. Belloni). Ces paramètres à l'échelle locale, forme et interactions, peuvent être étudiés en fonction de divers stimuli externes: température, pression, pH, champ magnétique, électrique, cisaillement etc... de façon à pouvoir décrire et comprendre les propriétés macroscopiques des systèmes.

**Références**

[1] *La Juste Argile*, éditeurs C. Williams, M. Daoud, Les Editions de Physique, **1995**. Un excellent ouvrage en langue française présentant les différents aspects de la matière molle !
[2] D. F. Evans, H. Wennerström, *The Colloidal Domain: Where Physics, Chemistry, Biology, and Technology Meet*, deuxième édition, Wiley, **1999**.
[3] J. Lyklema, *Fundamentals of Interface and Colloid Science*, Elsevier, **1991**.
[4] J. Israelachvili, *Intermolecular and Surface Forces*, Academic Press, New York, **1992**.
[5] P.G. De Gennes, *Scaling Concepts in Polymer Physics*; Cornell University Press: Ithaca, NY, **1979**.
[6] *Demain la Physique*, chapitre 7, sous la direction d'E. Brézin, Odile Jacob, **2004**.
[7] http://pubs.rsc.org/en/Journals/JournalIssues/SM
[8] C. Tanford C. J. Phys. Chem. 1972, 76, 3020-3024
[9] J. Israelachvili, D.J. Mitchell, B.W. Ninham, J. Chem. Soc., Faraday Trans II (1976), 72, 1525-1568.